\documentclass[sigconf]{acmart}

\AtBeginDocument{%
  \providecommand\BibTeX{{%
    \normalfont B\kern-0.5em{\scshape i\kern-0.25em b}\kern-0.8em\TeX}}}

\setcopyright{acmlicensed}
\copyrightyear{2024}
\acmYear{2024}
\acmDOI{XXXXXXX.XXXXXXX}

\acmConference[Conference acronym 'XX]{Make sure to enter the correct
  conference title from your rights confirmation email}{June 03--05,
  2018}{Woodstock, NY}
\acmBooktitle{Woodstock '18: ACM Symposium on Neural Gaze Detection,
 June 03--05, 2018, Woodstock, NY} 
\acmISBN{978-1-4503-XXXX-X/18/06}

\usepackage[whole]{bxcjkjatype}
\usepackage[unicode]{hyperref}

\usepackage{xcolor}
\usepackage{graphicx}

\usepackage{amsmath}
\usepackage{caption}
\usepackage{comment}
\usepackage{docmute}
\usepackage{empheq}
\usepackage{enumitem}
\usepackage{here}
\usepackage{karnaugh-map}
\usepackage{lscape}
\usepackage{multirow}
\usepackage{setspace}
\usepackage{siunitx}
\usepackage{url}
\usepackage{pbalance}

\usepackage{twemojis}
\usepackage{array}
\usepackage{amssymb} 
\usepackage{pifont}  
\usepackage{xcolor}  
\usepackage{algorithm}
\usepackage{algorithmic}
\usepackage{booktabs}
\usepackage{subcaption}
\usepackage[english]{babel}  
\usepackage{graphicx}  

\definecolor{ForestGreen}{RGB}{34,139,34}

\copyrightyear{2025}
\acmYear{2025}
\setcopyright{cc}
\setcctype{by}
\acmConference[ETRA '25]{2025 Symposium on Eye Tracking Research and Applications}{May 26--29, 2025}{Tokyo, Japan}
\acmBooktitle{2025 Symposium on Eye Tracking Research and Applications (ETRA '25), May 26--29, 2025, Tokyo, Japan}\acmDOI{10.1145/3715669.3726848}
\acmISBN{979-8-4007-1487-0/2025/05}

\begin{document}

\title{See-Through Face Display for DHH People: Enhancing Gaze Awareness in Remote Sign Language Conversations with Camera-Behind Displays}

\author{Kazuya Izumi}
\authornote{Both authors contributed equally to the paper}
\email{izumin@digitalnature.slis.tsukuba.ac.jp}
\affiliation{%
  \institution{University of Tsukuba}
  \city{Tsukuba}
  \country{Japan}
}

\author{Akihisa Shitara}
\authornotemark[1]
\orcid{0000-0002-8944-0023}
\email{
theta-akihisa@digitalnature.slis.tsukuba.ac.jp}
\affiliation{%
  \institution{University of Tsukuba}
  \city{Tsukuba}
  \country{Japan}
}



\author{Yoichi Ochiai}
\email{wizard@slis.tsukuba.ac.jp}
\affiliation{%
  \institution{\mbox{R\&D Center for Digital Nature}}
  \country{Japan}
}


\begin{abstract}
This paper presents a sign language conversation system based on the See-Through Face Display to address the challenge of maintaining eye contact in remote sign language interactions. A camera positioned behind a transparent display allows users to look at the face of their conversation partner while appearing to maintain direct eye contact. Unlike conventional methods that rely on software-based gaze correction or large-scale half-mirror setups, this design reduces visual distortions and simplifies installation. We implemented and evaluated a videoconferencing system that integrates See-Through Face Display, comparing it to traditional videoconferencing methods. We explore its potential applications for Deaf and Hard of Hearing (DHH), including multi-party sign language conversations, corpus collection, remote interpretation, and AI-driven sign language avatars. Collaboration with DHH communities will be key to refining the system for real-world use and ensuring its practical deployment.
\end{abstract}

\ccsdesc[500]{Human-centered computing~Displays and imagers}

\keywords{Remote Communication, Eye Contact, Eye Tracking, Sign Language, Deaf and Hard of Hearing, Avatar-Based Communication, Assistive technologies, Gaze-informed interfaces}

\begin{teaserfigure}
  \includegraphics[width=\textwidth]{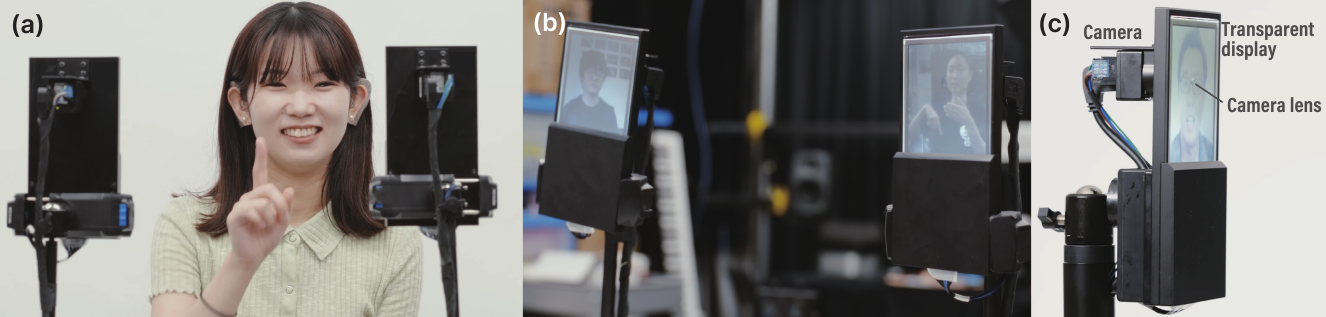}
  \caption{(a, b) Deaf and hard of hearing users communicate via sign language with a conversation partner displayed on the screen. (c) The See-Through Face Display consists of a transparent display and a camera placed directly behind it, so that the user's face appears superimposed on the camera lens (reproduced from \cite{Izumi2024-kn}).}
  \label{fig:teaser}
\end{teaserfigure}



\maketitle

\section{Introduction}

This position paper presents a discussion between the first author, who possesses expertise in gaze analysis and gaze interaction, and the second author, who is a Deaf person and has been conducting accessibility research for Deaf and hard of hearing (DHH). The paper focuses on accessibility applications that utilize gaze analysis and gaze interaction technologies for DHH.
The paper discusses applications that utilize See-Through Face Display technology by Izumi et al~\cite{Izumi2024-kn}, because we aim to the application incorporated a gaze natural interactions in sign language conversation as future work's goal.
The application examples include applications for online sign language conversations among multiple Deaf people, AI-powered avatar assistance, and remote sign language interpretation services for Deaf in settings such as reception areas.


Many studies on gaze analysis targeting DHH have conducted research, as reported in the survey paper by Agrawal and Peiris~\cite{ISeeWhat}.
Specifically, in sign language conversation, gaze shifts can mean indicators of starting to speak after a speaker changes~\cite{DeafEye-TalkTimming, Turn-talking-deaf}.
Furthermore, gaze interaction should be considered in UI design approaches such as ``Sign Language User Interface (SUI)''~\cite{WS-SUI}.
However, current UIs and systems often fail to consider gaze interaction needs of the Deaf.
This is likewise the case in display-based sign language conversation contexts, such as remote sign language interactions and sign language avatars, where the lack of direct eye contact remains a persistent challenge.

On the other hand, early attempts to support eye contact in remote conversations date back to systems such as ClearBoard~\cite{Ishii1992-gq}, and this functionality has since been integrated into commercial products such as NVIDIA Maxine Eye Contact\footnote{\url{https://developer.nvidia.com/blog/improve-human-connection-in-video-conferences-with-nvidia-maxine-eye-contact/}}.
Unlike these approaches, the display setup proposed by Izumi et al. features a smaller form factor, eliminates visual artifacts, and avoids image-processing techniques that can distort users' gaze.
This preservation of natural gaze relationships is particularly important in sign language conversation, where gaze cues play a critical role in turn-taking and engagement.
Given these advantages, we propose that this system is well-suited for remote video-based sign language conversation and have accordingly developed a multi-party videoconferencing system based on the See-Through Face Display.

In this paper, we introduce the developed system and present a comparison of user eye contact differences between our system and conventional videoconferencing systems.
Furthermore, we discuss evaluations focusing on the gaze behavior and user experience of DHH users, as well as future applications such as dataset collection of gaze and sign language interactions and the potential utilization of AI-driven sign language avatars.
\section{Related Work}

\subsection{Eye Contact in Remote Video Conversations}

Mutual gaze, or eye contact, is a key element of face-to-face communication and serves as an important channel for nonverbal signaling, shaping various aspects of social interaction~\cite{PhysiologicalAspectsCommunicationViaMutualGaze-Mazur1980-cu}.
In the context of remote video communication, however, the issue of lack of eye contact caused by the misalignment of the camera with a user's on-screen face has been a subject of discussion in the Human-Computer Interaction community for some time~\cite{Jaklic2017-dm}. To address this issue, two major approaches have been proposed: hardware-based and software-based solutions.

Notable hardware approaches employ half-mirrors or carefully positioned cameras to minimize parallax and achieve near-direct eye contact~\cite{Ishii1992-gq, Okada1994-oa, Otsuka2016-wx}.
These methods preserve natural gaze cues with minimal visual distortion. However, these methods require specialized equipment, involve complex setups, and often lack scalability for everyday use.
Software-based approaches include switching the video viewpoint based on head pose estimation \cite{Yang2002-uy,Vertegaal2003-wn} and manipulating eye regions in the video through image processing \cite{Wood2018-bq,wang2021facevid2vid,He2021-ps}. While these methods do not require the complex setups characteristic of hardware-based approaches, these methods often introduce visual artifacts, such as misaligned eye regions or jittery gaze movements, which can reduce perceived authenticity and engagement.


The See-Through Face Display bridges the gap between hardware and software solutions by integrating a transparent display and an embedded camera into a compact, user-friendly design.
Unlike traditional hardware-based approaches, it eliminates complex installations, and unlike software-based methods, it avoids noticeable visual distortions.
The display determines each user's position relative to others. This setup ensures that when two participants make eye contact, it appears to outside viewers that only these two individuals are exchanging glances.

\subsection{Target to Deaf and Hard of Hearing}
Regarding gaze analysis for DHH, multiple studies have reported and there is a summarized survey paper~\cite{ISeeWhat} that compiles these findings.
In general, gaze related elements are sometimes used as grammatical components in sign language~\cite{Sign-Gaze}. 
In addition, gaze plays an important role during speaker transitions and the initiation of utterances \cite{DeafEye-TalkTimming, Turn-talking-deaf}.
Thompson et al. also reported relationships between eye gaze and verb agreement in sign language. ~\cite{EyeGeze-deaf-sign}.  
However, issues with gaze interaction for DHH have been noted, including the mismatch between gaze direction and spatial position in online environments~\cite{OnlineEnv-DHH}.
On the other hand, some cases include avatars for systems and services, such as a kiosk service system and a personal assistant, etc., to make Deaf people usable.
However, to our knowledge, few studies have addressed gaze interaction in conversations with virtual characters for DHH users.
For example, in a study on sign language avatars~\cite{SigningAvatar-Eye}, the following comments were provided based on participant feedback:

\begin{quotation}
  Permanent eye contact was regarded as unnatural and causing discomfort.
\end{quotation}

From this comment, there is a need to discuss design approaches that facilitate natural gaze interaction in sign language conversations between Deaf people. Specifically, Deaf people can have gaze interactions that allow gaze shifts that can make eye contact or look away, as well as real-life interactions.

\section{System}\label{sec:system}

\begin{figure}[htbp]
  \centering
  \includegraphics[width=\linewidth]{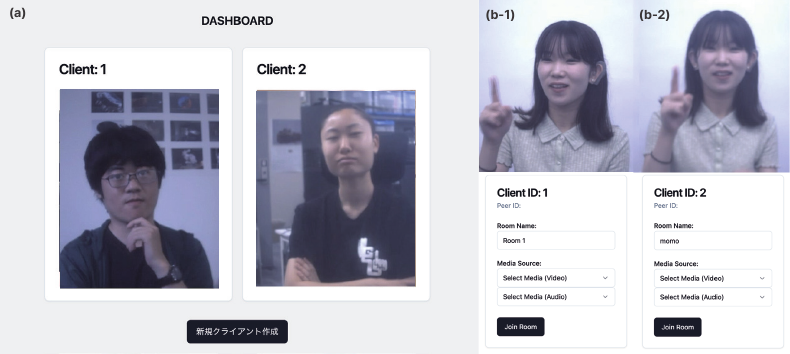}
  \caption{The screenshots of our videoconferencing software. (a) The user accesses the dashboard page to establish one client connection per conversational partner. (b) The client interface as it appears on the See-Through Face Display. For debugging purposes, the user can access the shared room, as well as control the camera and microphone on the See-Through Face Display. Since the display renders only the top-left 320$\times$360 pixels of the screen, it exclusively shows the remote user's face.}
  \label{fig:user_interface}
\end{figure}

\begin{figure}[htbp]
  \centering
  \includegraphics[width=\linewidth]{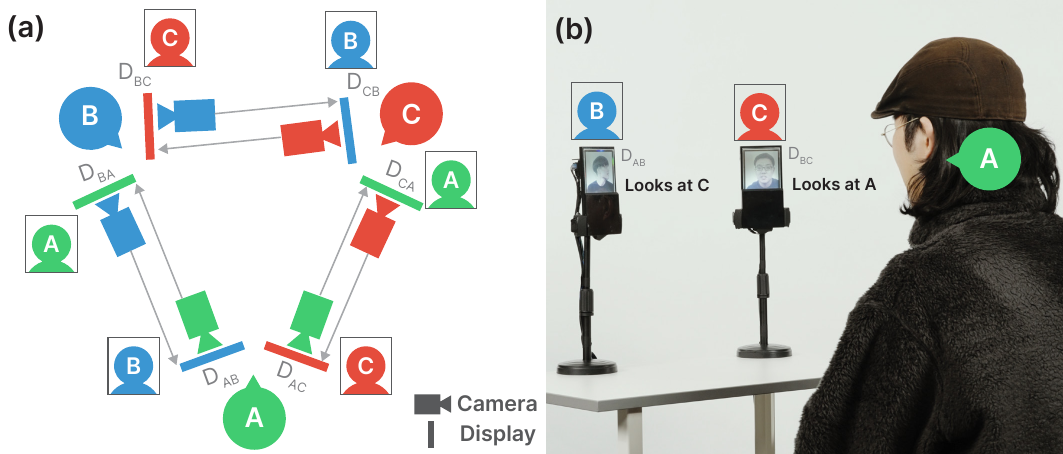}
  \caption{(a) Example of a P2P mesh communication configuration in a three-party conversation.
  User A establishes P2P connections with Users B and C via Displays $D_{AB}$ and $D_{AC}$, respectively. Displays $D_{AB}$ and $D_{AC}$ show images of Users B and C captured by cameras positioned behind $D_{BA}$ and $D_{CA}$, respectively, while Displays $D_{BA}$ and $D_{CA}$ show images of User A captured by cameras behind $D_{AB}$ and $D_{AC}$.  
  (b) User A is speaking; Users B and C are displayed separately in $D_{AB}$ and $D_{AC}$ and can make eye contact with each other.}
  \label{fig:system_design}
\end{figure}

See-Through Face Display synchronizes a transparent display with a camera positioned behind it, allowing the user's face to be captured without obstructing the displayed content~\cite{Izumi2024-kn}. 
During use in the remote conversations, the face of the interlocutor appears on the transparent display, ensuring that when the user looks at this face, the transmitted video portrays them as looking straight ahead.
We have developed videoconferencing software that utilizes this display to facilitate face-to-face like remote sign language conversation among two or more participants by preserving gaze alignment.
The user interface of the software is shown in \autoref{fig:user_interface}.
Each interlocutor is displayed on a separate See-Through Face Display, which the user arranges in an arc in front of them, as depicted in \autoref{fig:system_design}(a).
Using the software's dashboard, the user sets up a separate client for each conversation partner, enabling gaze-aligned remote sign language conversation via WebRTC-based peer-to-peer (P2P) connections.
For example, in a three-party remote sign language conversation, the system constructs a P2P mesh network, as illustrated in \autoref{fig:system_design}(b).

\begin{figure}[htbp]
  \centering
  \includegraphics[width=\linewidth]{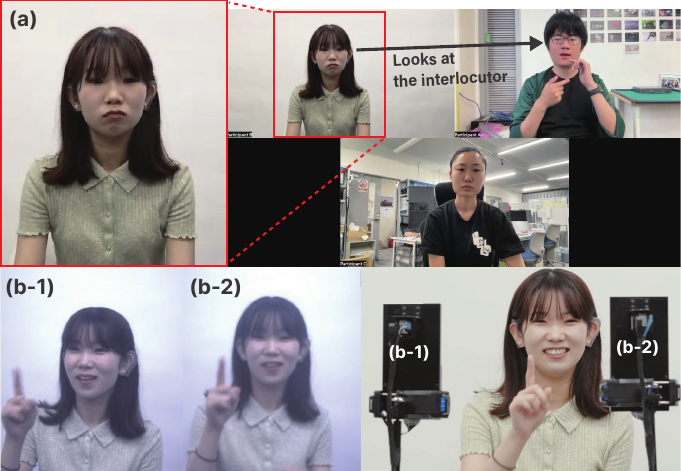}
  \caption{(a) In a typical videoconferencing setup, the user is in fact looking at their conversation partner, but the transmitted image depicts them as if they are looking away. 
(b-1) In See-Through Face Display, for any conversation partner the user is not looking at, the video stream appears as though the user is looking elsewhere.
(b-2) Conversely, for the partner the user is actually looking at, the video stream appears as though the user is looking directly at them.}
  \label{fig:result}
\end{figure}

\autoref{fig:result} compares three-party sign language conversations using a conventional videoconferencing system (e.g., Zoom\footnote{\url{https://www.zoom.com/}}) with those using the See-Through Face Display.
With the See-Through Face Display, participants displayed on each other's screens can maintain genuine eye contact throughout the interaction.

However, as shown in \autoref{fig:result}(b), the display has a lower resolution than conventional monitors and exhibits more pronounced flicker, which may affect the clarity of facial expressions and hand movements essential for sign language conversation.
High-quality video is critical for accurately understanding signs and improving the overall conversational experience~\cite{Perception-Sign, MobileASL}.
These issues stem from technical constraints, including the need to synchronize the camera's shutter speed with the display's refresh rate, and the limitations of the display's internal driving frequency.
We plan to address these challenges in future prototypes.
Additionally, the current prototype does not provide immediate feedback to users on whether their faces or signing are being properly captured, requiring them to periodically check their video feed on the dashboard during conversations. An important future challenge is designing an unobtrusive feedback mechanism that alerts users when their face or signing is not properly displayed, without disrupting the flow of conversation.

\section{Future Application Idea}


When applying the system described in Section~\ref{sec:system} to applications designed for DHH, we developed multiple ideas. Among these, some are inspired by SUI~\cite{WS-SUI}.

\subsection{Online Environment Conversation}
To verify whether this contributes to solving the challenges in online conversational environments~\cite{OnlineEnv-DHH}, we plan to conduct gaze analysis comparing our system with Zoom as an example of an online environment.
We observe the quality of eye contact and changes in communication by examining how participants look at each conversation partner's face in different contexts, and identify which aspects of gaze present challenges.
Based on these findings, it becomes possible to develop guidelines for conversational systems in online environments tailored to DHH individuals.

\subsection{Sign Language Data Corpus}
Several instances of building a sign language dataset have been reported and made publicly available, involving not only the information that the method of collecting sign language videos and studio setups for filming~\cite{Auslan-Daily, RWTH-PHOENIX-Weather, ms-asl, How2Sign, MM-WLAuslan, GSL-DataSet}.
Furthermore, there are reported cases of datasets built from publicly available videos uploaded to YouTube ~\cite{YouTube-ASL, YouTube-SL-25}.
There is also research correcting camera angles to a frontal position to provide linguists direct and human-readable access to the collecting sign language data~\cite{skobov-bono-2023-making}.
However, examples of collecting sign language data and gaze data as a paired set appear to be lacking, like collecting paired speech data and gaze data of hearing people~\cite{EyeMovementControlledDialogueSetting-Dembinsky2024-lo} and so there can’t synthesis natural eye gaze during sign language without the paired dataset.
Thus, using our system, it becomes possible to collect raw gaze data without the need to generate from sign language video, potentially expanding approaches for new sign language data corpus methodologies.
Nevertheless, it is necessary to systematically evaluate the advantages and disadvantages of this approach compared to alternatives, such as face-to-face multi-person conversations or approaches using both glasses-type eye trackers and motion capture.


\subsection{Sign Language Conversation with Human or AI}

\begin{figure}[htbp]
  \centering
  \includegraphics[width=\linewidth]{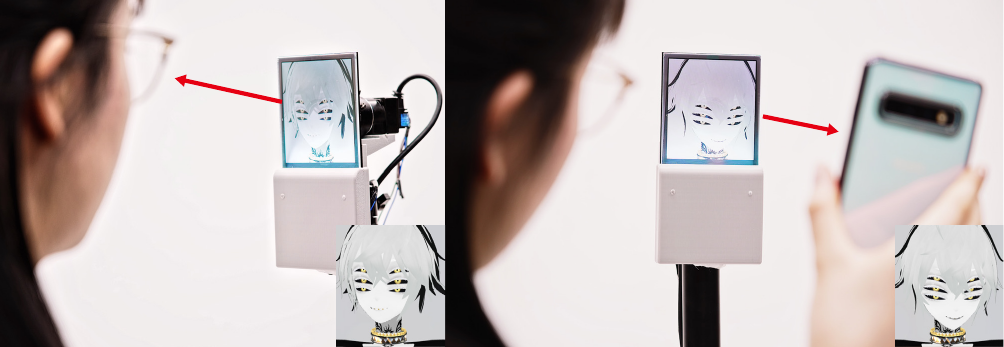}
  \caption{Implementation of eye contact communication with AI avatars by Izumi et al. (reproduced from \cite{izumi2025animegazerealtimemutualgaze}). \copyright ZEPTO002, licensed under VN3 license version 1.10.}
  \label{fig:eye_contact}
\end{figure}

Gaze alignment issues, which have been identified as a challenge in remote sign language interpretation settings~\cite{RemoteSignInterpreter}, may be addressed by our system.
This enables sign language interpreters to perform several tasks more effectively: (1) determine whether the Deaf person is looking at them or at individuals at the physical location, and (2) express spatial relationships more naturally, supported by improved shared spatial awareness.
The first example is that the sign language interpreter easily determines whether a Deaf person is looking at the clerk or the product when they talk to the clerk through remote sign language interpretation in case shopping. 
The second example is that the sign language interpreter can easily express ``Walk'' verb and pointing to places the Deaf person wants to go through interpretation directions from the store clerk.
This would improve accessibility when using remote sign language interpretation services at various customer service points such as stores and government offices.
Research on sign language avatars has been conducted not exclusively for conversations with humans, but equivalently for conversations with AI, generating sign language animations through a sequential process of converting text to sign language sequences and subsequently to motion~\cite{SignMotionGeneration}.
Furthermore, the virtual human, KIKI~\footnote{NHK ENTERPRISES, Photorealistic digital avatar ``KIKI'' \url{https://www.nhk-ep.co.jp/signlanguage/en/}}, developed by NHK ENTERPRISES, is also cited as a representative example of digital humans used for sign language avatars. This suggests that if digital humans and other sign language avatars could be deployed not only for remote sign language interpretation but also for customer service interactions, incorporating gaze interaction features would likely reduce stress for DHH users who spoke sign language and facilitate more efficient sign language conversations.
Multiple studies have reported accessibility analysis for personal assistants designed for DHH~\cite{A11y-CUI, Analay-CUI}. In particular, research on sign language based personal assistants~\cite{SUI-VUI} mentions that sign language considerations should include gaze elements. In addition, there are cases investigating gaze as one of the wake-up methods for personal assistants~\cite{kato-SUI-2021, kato-SUI-2022}. Therefore, in terms of utilizing gaze interaction for conversations with personal assistants, this could become one of the applications of our system.

In recent years, gaze synthesis techniques for AI voice interaction avatars displayed on screens have been actively discussed~\cite{Real-timeConversationalGazeSynthesisAvatars-Canales2023-sp, EyeMovementControlledDialogueSetting-Dembinsky2024-lo}.
Izumi et al. have also conducted evaluations using the See-Through Face Display to enhance the gaze awareness of AI avatars, as shown in \autoref{fig:eye_contact}~\cite{izumi2025animegazerealtimemutualgaze}.
Building on this, discussing gaze synthesis techniques tailored for sign language conversation is crucial for improving the accessibility of AI-based conversation in the future.
However, since the current prototype of the See-Through Face Display measures only about four inches, it is difficult to simultaneously show both the avatar's face and hands.
Therefore, we plan to develop a larger prototype capable of showing an avatar's upper body in the future.
\section{Conclusion}
This paper examined the See-Through Face Display system, which enhances gaze communication on remote conversations, and its potential applications in remote sign language conversation. Enhancing eye contact with the See-Through Face Display can improve DHH conversation in an online 
environment. Additionally, natural interactions that incorporate gaze alignment may increase user immersion and facilitate smoother information exchange in sign language corpus collection, remote interpretation, and AI-driven sign language avatars.

Future work will focus on improving display resolution, reducing flicker, and developing user-friendly tools for video verification. We also plan to assess the system's effectiveness through real-world testing with DHH users, analyzing their gaze patterns. As noted in \cite{InterdisciplinaryPerspective}, close collaboration with DHH communities is essential. We will continue development in close collaboration with stakeholders and DHH communities to ensure the system meets real-world needs.




\begin{acks}
We are grateful to Japan Display Inc. for lending us the prototype of See-Through Face Display and Kazuhiko Sako, Kazunari Tomizawa, and Kentaro Okuyama for their technical assistance in the hardware development.
\end{acks}

\bibliographystyle{ACM-Reference-Format}
\bibliography{reference}

\end{document}